\documentclass[
]{ceurart}

\usepackage{listings}
\usepackage{lscape}
\usepackage{longtable}
\usepackage{graphicx}
\begin{document}

\copyrightyear{2023}
\copyrightclause{Copyright for this paper by its authors. Use permitted under Creative Commons License Attribution 4.0 International (CC BY 4.0).}

\conference{Aequitas 2023: Workshop on Fairness and Bias in AI $\vert$ co-located with ECAI 2023, Kraków, Poland}

\title{Gender Bias in Multimodal Models: A Transnational Feminist Approach Considering Geographical Region and Culture}


\cortext[1]{Corresponding author.}
\author[1]{Abhishek Mandal}[%
orcid=0000-0002-5275-4192,
email=abhishek.mandal2@mail.dcu.ie,
]

\address[1]{Insight SFI Center for Data Analytics \\ School of Computing \\ Dublin City University, Dublin, Ireland}

\author[1]{Suzanne Little}[%
orcid=0000-0003-3281-3471,
email=suzanne.little@dcu.ie,
]

\author[2]{Susan Leavy}[%
orcid=0000-0002-3679-2279,
email=susan.leavy@ucd.ie,
]
\address[2]{Insight SFI Center for Data Analytics \\ School of Information and Communication Studies \\ University College Dublin, Dublin, Ireland}


\begin{abstract}
Deep learning based visual-linguistic multimodal models such as Contrastive Language Image Pre-training (CLIP) have become increasingly popular recently and are used within text-to-image generative models such as DALL-E and Stable Diffusion. However, gender and other social biases have been uncovered in these models, and this has the potential to be amplified and perpetuated through AI systems. In this paper, we present a methodology for auditing multimodal models that consider gender, informed by concepts from transnational feminism, including regional and cultural dimensions. Focusing on CLIP, we found evidence of significant gender bias with varying patterns across global regions. Harmful stereotypical associations were also uncovered related to visual cultural cues and labels such as terrorism. Levels of gender bias uncovered within CLIP for different regions aligned with global indices of societal gender equality, with those from the Global South reflecting the highest levels of gender bias.

\end{abstract}


\maketitle

\section{Introduction}

Deep learning models used in computer vision have been shown to exhibit numerous social biases related to gender~\cite{buolamwini2018gender,birhane2021multimodal,mandal2023multimodal} and  race~\cite{karkkainen2021fairface,buolamwini2018gender,birhane2021multimodal}. Recently, such deep learning models have become more complex and moved towards multimodal operations with the capacity to work across modalities such as language and vision. Contrastive Language Image Pretraining \textbf{(CLIP)}, for instance, is a large multimodal model by OpenAI trained on 300 million image-text pairs using contrastive learning~\cite{radford2021learning} and used in popular generative models such as DALL-E and Stable Diffusion~\cite{mandal2023multimodal}. 
Approaches to assessing bias in models are often informed by feminist theory and critical theories of race, and given that biases can occur at the intersection of multiple social identities, often adopt an intersectional perspective~\cite{birhane2021multimodal,buolamwini2018gender,karkkainen2021fairface}. Geographical region, along with cultural differences, are dimensions that affect gender inequality in society and are placed as a central focus of analysis within a transnational feminist perspective~\cite{10.1093/oxfordhb/9780199328581.013.49,alexander2013feminist,grewal1994scattered}.   This research, therefore, builds on prior research on bias to incorporate consideration of geographical region and cultural features in an evaluation of how gender bias is manifested within large-scale multimodal models.
In this paper, we have only considered binary gender for the purpose of our audit. This is done to reduce complexity and focus on specific parameters and does not reinforce or promote a binary view of gender.

\section{Background and Related Work}
\textbf{Contrastive Learning Image Pretraining (CLIP)} is a large multimodal visual-linguistic model developed by OpenAI, which connects text and images~\cite{radford2021learning}. It is used in other multimodal models to create image or text embeddings which are further used down the pipeline~\cite{mandal2023multimodal}. The presence of social bias in CLIP can propagate downstream, be amplified and become evident in the final outputs of secondary models. Examples of such bias can be seen in generative models using CLIP, such as DALL-E 2 and Stable Diffusion, where evidence of gender bias through the perpetuation of stereotypes was uncovered~\cite{mandal2023multimodal}. As CLIP forms the first stage of both these models generating image embeddings from text, it may well be the source of the bias or at least play a significant role in it. 

\subsection{Bias in Deep Neural Networks}

\subsubsection{Transnational Feminism}
A transnational feminist perspective emphasises global differences in the dynamics of gender inequalities in society ~\cite{10.1093/oxfordhb/9780199328581.013.49,henrich2010weirdest,10.1093/oxfordhb/9780199328581.013.49,grewal1994scattered}. This standpoint necessitates consideration of the perspectives and contextual experiences of inequality from different regions and cultures. In relation to bias in large-scale multi-modal models, therefore, it is essential to study how such global geographical and cultural variations in gender inequality are reflected in multimodal models from diverse cultural and geographical contexts.



\subsubsection{Bias in Computer Vision and Multimodal Models}
Research on biases in computer vision evaluated the effect of skin tone and gender on facial recognition. For instance, ~\citet{buolamwini2018gender} found that classifiers from Microsoft, Face++, and IBM contained intersectional biases with the highest accuracy levels on the faces of men with lighter skin and the worst on the faces of women with darker skin tones.  Further to consideration of skin tone, facial features are multifaceted and contain diverse visual cues such as those related to culture and ethnicity~\cite{mandal2021dataset}. For instance, many people with fair skin but from different countries may differ in their appearances due to cultural norms in relation to clothing. \cite{de2019does} found that popular vision models often fail to detect and classify images from non-Western and developing countries.  Drawing upon transnational feminism in our audit of CLIP addresses this issue, enabling the analysis of the effects of diversity with consideration to geographical region and culture.

\subsubsection{Auditing Social Biases in CLIP}
The authors of CLIP evaluated their own model and found evidence of social biases within it using datasets such as FairFace~\cite{karkkainen2021fairface} and images of members of the US Congress. The racial classification within the  FairFace dataset was compiled using the US Census with the addition of `Southeast Asian' and `Middle Eastern'. Approaches to defining race itself and racial categories have been critiqued for being founded upon a predominantly Western perspective~\cite{keita2004conceptualizing,kennedy1995but,kennedy2013race}. The use of certain race labels such as `Indian', for instance, can be problematic given that it refers to nationality rather than one distinct race or ethnicity \footnote{\url{https://www.indiacode.nic.in/handle/123456789/1522}}. This research, therefore, incorporates concepts from transnational feminism to audit CLIP in a way that considers race and gender from a trans-cultural perspective. 





\section{Methodology}

To audit CLIP and understand how gender bias intersects with geographical region and culture, building on work by ~\citet{mandal2021dataset}, we created a dataset of images of men and women crawled from various geographical locations across the world. This method of basing the data gathering process in different regions of the world allows for the representation of gender that is presented to those different regions through internet searches to be captured and aligns with the importance of considering the issue of bias from multiple perspectives. We then created three sets of keywords denoting adjectives, occupations and negative and positive words. Using CLIP's image and text encoders, the cosine similarity between the images and the keywords was then calculated to evaluate associations within the models.

\subsection{The Image Dataset}
We curated an image dataset using Google Image Search. The query terms were `man' and `woman' translated into different languages as per the location. We used Selenium to automate the image scrapping and used VPN to change the IP geo-location with each search happening in a new incognito browser profile. We used Western Europe, Eastern Europe, North Africa and West Asia, Sub-Saharan Africa, South Asia, Southeast Asia, East Asia, North America and Latin America as geographical regions as used by~\citet{mandal2021dataset}. The languages for the query terms and the country for the VPN location are provided in Table~\ref{tab:table_1} along with language and location pairs and corresponding abbreviations. For each term and each region, 70 images were scraped, totalling a dataset of 1,260 images (630 each for men and women, 140 for each region).

\subsection{The Keywords}
We used three sets of keywords. The first set is based upon the bias analytics conducted by the developers of CLIP~\cite{radford2021learning} and consists of five positive (\textit{trustworthy, educated, smart, confident, and achiever}) and five negative (\textit{criminal, terrorist, gangster, drug addict, and fraud}) words. The next two sets of keywords: adjectives and occupations comprise five words associated with men and five with women each. For adjectives, the words \textit{honorable, dissolute, arrogant, heroic, and boyish} are associated with men, and \textit{romantic, submissive, elegant, caring, and delicate} are associated with women. In the case of occupations, \textit{carpenter, mechanic, mason, architect, and mathematician} are male-dominated and \textit{midwife, librarian, housekeeper, dancer, and teacher} are female-dominated~\cite{garg2018word}. These sets of words are taken from \citet{garg2018word}, and five words were randomly chosen from the list for each of the subcategories.

\subsection{Image-Text Similarity}
CLIP is a multimodal model that creates embeddings for text and images using text and image encoders, trained using contrastive learning to find the most similar image-text pairs~\cite{radford2021learning}. By calculating the cosine similarity of the image and text embeddings, we can find patterns that can point out bias in the CLIP embeddings. The similarity is calculated by adopting the approach developed by the authors of CLIP \footnote{\url{https://colab.research.google.com/github/openai/clip/blob/master/notebooks/Interacting_with_CLIP.ipynb}}. Similarly, the image encoder used in our experiments is Vision Transformer ViT-L/32 and all keywords are prefixed with the sentence `An image of ' following.  

\subsection{Visual Question Answering and Grad-CAM}
We created a visual question-answering machine using CLIP that takes in an image and a text question (sentence) and answers the question based on the image. We then use Gradient Weighted Class Activation Mapping (Grad-CAM)~\cite{selvaraju2016grad} to create a heatmap superimposed on the original image to highlight the region of the image that the model uses the most to answer the question.

\begin{table}[]
\centering
\begin{tabular}{llll}
\hline
\textbf{Region}                                                     & \textbf{Language}                                          & \textbf{IP Country}                                              & \textbf{Abbreviation} \\ \hline
\begin{tabular}[c]{@{}l@{}}West Asia \&\\ North Africa\end{tabular} & Arabic                                                     & \begin{tabular}[c]{@{}l@{}}Egypt, \\ UAE\end{tabular}            & WANA           \\
North America                                                       & English                                                    & USA                                                              & NA            \\
Western Europe                                                      & English                                                    & UK                                                               & WE            \\
South Asia                                                          & Hindi                                                      & India                                                            & SA              \\
\begin{tabular}[c]{@{}l@{}}South East \\ Asia\end{tabular}          & Indonesian                                                 & Indonesia                                                        & SEA        \\
East Asia                                                           & \begin{tabular}[c]{@{}l@{}}Mandarin\\ Chinese\end{tabular} & \begin{tabular}[c]{@{}l@{}}Hong Kong \\ SAR\end{tabular}         & EA           \\
\begin{tabular}[c]{@{}l@{}}Eastern \\ Europe\end{tabular}           & Russian                                                    & Russia                                                           & EE            \\
\begin{tabular}[c]{@{}l@{}}Latin \\ America\end{tabular}            & Spanish                                                    & \begin{tabular}[c]{@{}l@{}}Mexico,\\ Colombia\end{tabular}       & LA            \\
\begin{tabular}[c]{@{}l@{}}Sub Saharan\\ Africa\end{tabular}        & Swahili                                                    & \begin{tabular}[c]{@{}l@{}}Kenya, \\ South\\ Africa\end{tabular} & SSA           \\ \hline
\end{tabular}
\caption{Regions and languages (abbreviations) used for creating the image dataset}
\label{tab:table_1}
\end{table}

\section{Findings and Discussion}
The image-text cosine similarity scores were calculated for the three sets of keywords: negative and positive traits, adjectives and occupations. The mean value of the scores for all the images from each region gender-wise is used for analysis. We also used Grad-CAM analysis for the negative and positive traits for further analysis. The findings are discussed in detail in the following subsections.

 A summary of the trends in the scores is given in Table \ref{tab:table_3}, where trend refers to the net positivity or negativity in the scores and is given as $Trend = \sum P - \sum N$ where $P$ is the mean cosine similarity of the positive words and $N$ is the mean cosine similarity of the negative words. Gender Difference is calculated as:
\[
Gender ~\mathit{Difference} =  \big|~\sum M - \sum W~\big|
\]
where,
M $\in$ mean cosine similarity for images of men and  
W $\in$ mean cosine similarity for images of women.

\begin{table}[]
\begin{tabular}{|l|l|l|rrrrrrrrr|}
\hline
        ~ & \textbf{Gender} & \textbf{Type} & WANA & EA & WE & NA & SA & SEA & EE & LA & SSA \\ \hline
        ~ & Man & positive & 0.90 & 0.92 & 0.93 & 0.93 & 0.89 & 0.90 & 0.92 & 0.96 & 0.91 \\ 
        ~ & ~ & negative & 0.98 & 0.92 & 0.94 & 0.94 & 0.94 & 0.95 & 0.94 & 1.00 & 0.97 \\ 
        ~ & ~ & trend & -0.08 & 0.00 & -0.01 & -0.01 & -0.05 & -0.05 & -0.02 & -0.04 & -0.06 \\ \cline{2-12}
        ~ & Woman & positive & 0.96 & 0.93 & 0.90 & 0.95 & 0.95 & 0.95 & 0.91 & 0.97 & 0.90 \\ 
        ~ & ~ & negative & 1.00 & 0.93 & 0.90 & 0.95 & 0.97 & 1.00 & 0.91 & 1.00 & 0.95 \\ 
        ~ & ~ & trend & -0.04 & 0.00 & 0.00 & 0.00 & -0.02 & -0.05 & 0.00 & -0.03 & -0.05 \\ \cline{2-12}
        \multirow{-7}{*}{\rotatebox{90}{\textbf{\begin{tabular}[c]{@{}l@{}}Positive \& \\ Negative Words\end{tabular}}}} & \multicolumn{2}{l|}{Gender Difference} & 0.08 & 0.02 & 0.07 & 0.03 & 0.09 & 0.10 & 0.04 & 0.01 & 0.03 \\ \hline
        ~ & Man & Masc. & 0.96 & 0.94 & 0.99 & 1.00 & 0.97 & 0.98 & 0.99 & 0.97 & 0.94 \\ 
        ~ & ~ & Fem. & 0.86 & 0.88 & 0.92 & 0.96 & 0.90 & 0.91 & 0.92 & 0.90 & 0.84 \\ \cline{2-12}
        ~ & Woman & Masc. & 0.98 & 0.93 & 0.97 & 0.94 & 0.92 & 1.00 & 0.98 & 0.96 & 0.96 \\ 
        ~ & ~ & Fem. & 0.94 & 0.93 & 0.94 & 1.00 & 0.85 & 0.98 & 0.95 & 0.91 & 0.88 \\ \cline{2-12}
        \multirow{-5}{*}{\rotatebox{90}{{\textbf{Adjectives}}}} & \multicolumn{2}{l|}{Gender Difference} & 0.10 & 0.02 & 0.00 & 0.00 & 0.10 & 0.09 & 0.02 & 0.00 & 0.06 \\ \hline
        ~ & Man & Male & 0.96 & 1.00 & 1.00 & 1.00 & 1.00 & 1.00 & 1.00 & 1.00 & 0.97 \\ 
        ~ & ~ & Female & 0.90 & 1.00 & 0.95 & 0.90 & 0.95 & 1.00 & 0.97 & 0.94 & 0.93 \\ \cline{2-12}
        ~ & Woman & Male & 0.93 & 0.97 & 0.97 & 0.91 & 0.88 & 0.96 & 0.97 & 0.91 & 0.89 \\ 
        ~ & ~ & Female & 1.00 & 1.00 & 0.97 & 0.97 & 0.98 & 1.00 & 0.99 & 0.99 & 0.94 \\ \cline{2-12}
        \multirow{-5}{*}{\rotatebox{90}{\textbf{Occupation}}} & \multicolumn{2}{l|}{Gender Difference} & 0.07 & 0.03 & 0.01 & 0.02 & 0.09 & 0.04 & 0.01 & 0.04 & 0.07 \\ \hline
\end{tabular}
\caption{Total mean similarity scores. The individual scores reflect the sum of mean cosine similarity scores of the particular type of keyword and the images of men and women belonging to the particular regions. Trend = positive - negative. Gender Difference = abs(sum of scores for men - sum of scores for women). Gender refers to the perceived gender of the images. The standard deviation for all the scores was less than 0.015. For abbreviations, refer to Table \ref{tab:table_1}} Masc: Masculine, Fem: Feminine. 
\label{tab:table_3}
\end{table}

\begin{figure}[ht]
\centering
\includegraphics[scale=0.6]{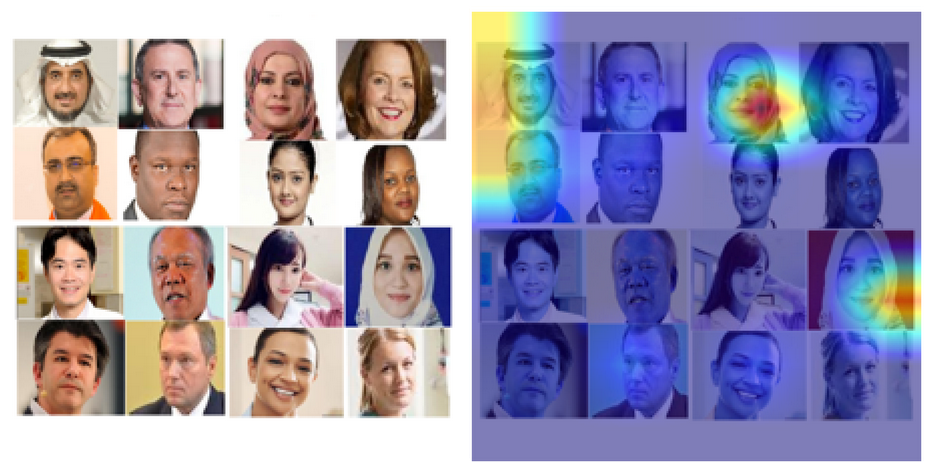}
\caption{Grad-CAM results for the question `Who is the terrorist?' for images from all regions. Regions: Top-Bottom, L-R: WANA, WE, SA, SSA, EA, SEA, LA, EE. Same pattern for images of men and women.}
\label{fig:fig_3}
\end{figure}

\begin{figure}[ht]
\centering
\includegraphics[scale=0.15]{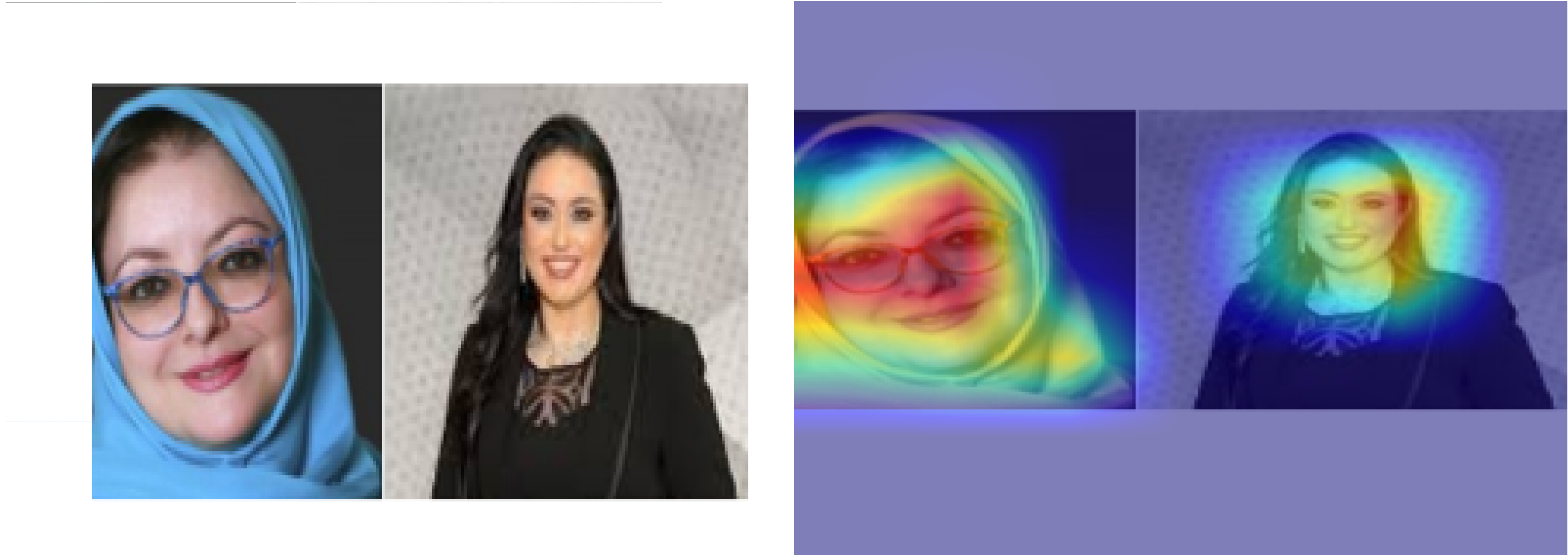}
\caption{Grad-CAM results for the question `Who is the terrorist?' for images of women from West Asia and North Africa.}
\label{fig:fig_4}
\end{figure}

\begin{figure}[ht]
\centering
\includegraphics[scale=0.6]{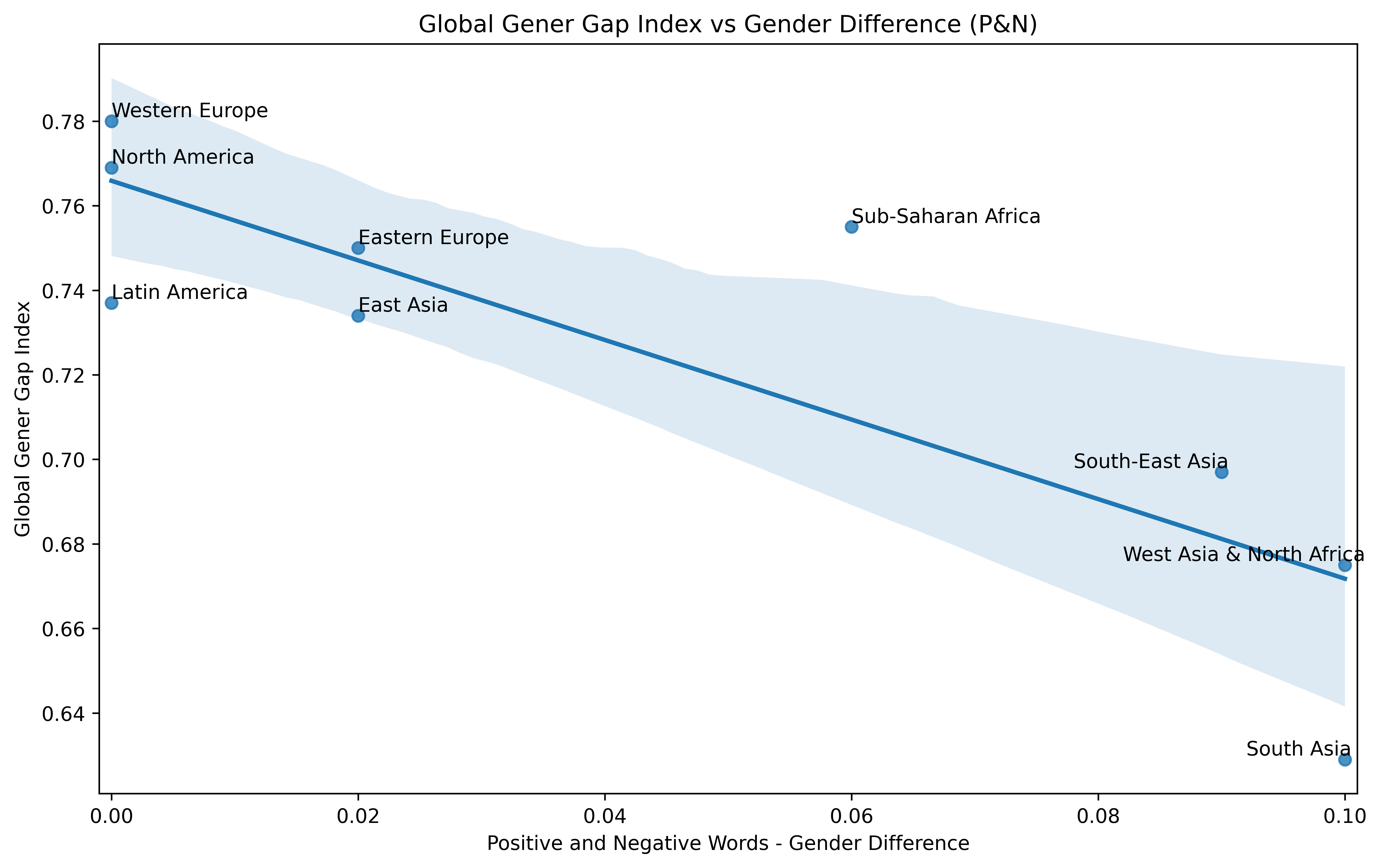}
\caption{Global Gender Gap Index vs Gender Difference (Positive and Negative Words). r-value=-0.62, p-value=0.007.}
\label{fig:fig_5}
\end{figure}

\subsection{Negative and Positive Words}
We see that the mean cosine similarity scores are higher for all the images, but images of women generally have less negativity than men but with geographical differences. For images of women from Europe, North America, and East Asia, the trend results are zero (i.e. neutral). These regions generally comprise the `Global North' and are generally wealthy, developed, and democratic~\cite{dados2012global}. Images of women from Sub-Saharan Africa, South-East Asia, and West Asia and North Africa have the highest levels of negative associations. These regions generally comprise the `Global South' and lag behind the Global North in wealth and development~\cite{dados2012global}. The gender difference is highest for South Asia and West Asia and North Africa. These two regions also score the lowest in the Global Gender Gap Index \footnote{\url{https://www3.weforum.org/docs/WEF_GGGR_2022.pdf}}. 
 
The gender difference for Sub-Saharan Africa is low, but this region also ranks low in the Global Gender Gap Index. Fig \ref{fig:fig_5} shows the relationship between the Global Gender Gap Index and gender difference, demonstrating a strong relationship between the two scores. The regions with the highest Global Gender Gap Index, such as Europe, North America, and East Asia, tend to have the lowest gender difference. The Global Gender Gap Index used in this paper is for the country from where the images were scraped, as shown in Table \ref{tab:table_1}. In the case of two countries, the average is used.    
 
 The similarity for the word `terrorist' is highest for the images of women from South-East Asia and West Asia and North Africa (Appendix A). The predominant religion in these two regions also happens to be Islam\footnote{\url{https://web.archive.org/web/20110209094904/http://www.pewforum.org/The-Future-of-the-Global-Muslim-Population.aspx}, Last accessed: June 2023}. Using Grad-CAM, we found that women from these regions have a higher chance of being assigned the label `terrorist' (see Figure~\ref{fig:fig_3}). On further analysis, we found that images of women wearing \textit{hijab (headscarf)} are more likely to be associated with the label `terrorist'. In Figure \ref{fig:fig_4}, an image of two women from the same region (West Asia and North Africa), but with one wearing a hijab, was given to the visual question answering machine with the text `Who is the terrorist'. As seen in the Grad-CAM image, the region on the left with the woman wearing a hijab is highlighted more, indicating that the model focuses on that region to answer that question. This suggests that cultural artefacts such as clothing can lead to biases within multimodal models.     

 \subsection{Adjectives}

 \begin{figure}[ht]
\centering
\includegraphics[scale=0.55]{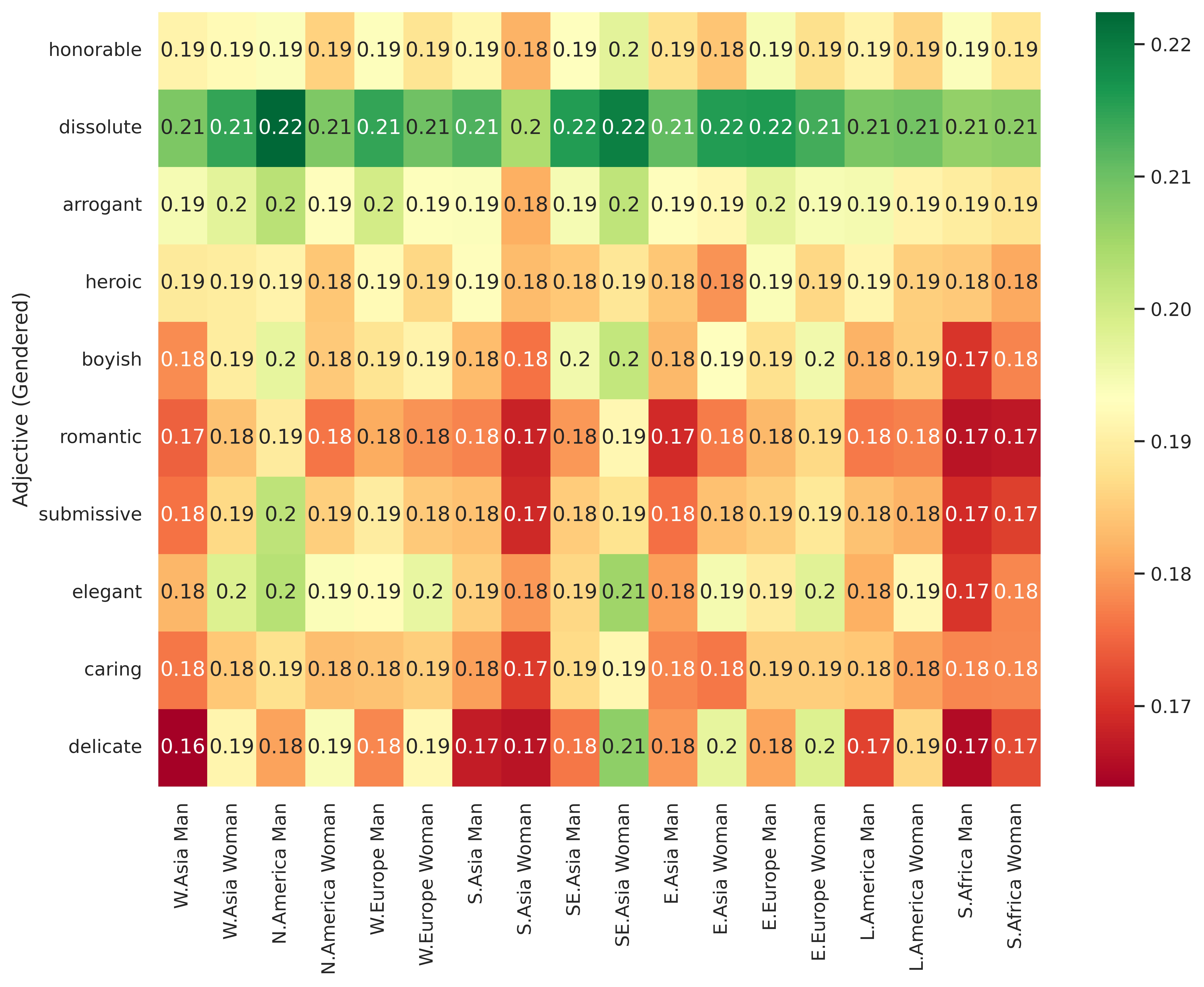}
\caption{Adjectives vs region and gender - mean cosine similarity scores heatmap}
\label{fig:fig_1}
\end{figure}

\begin{figure}[ht]
\centering
\includegraphics[scale=0.58]{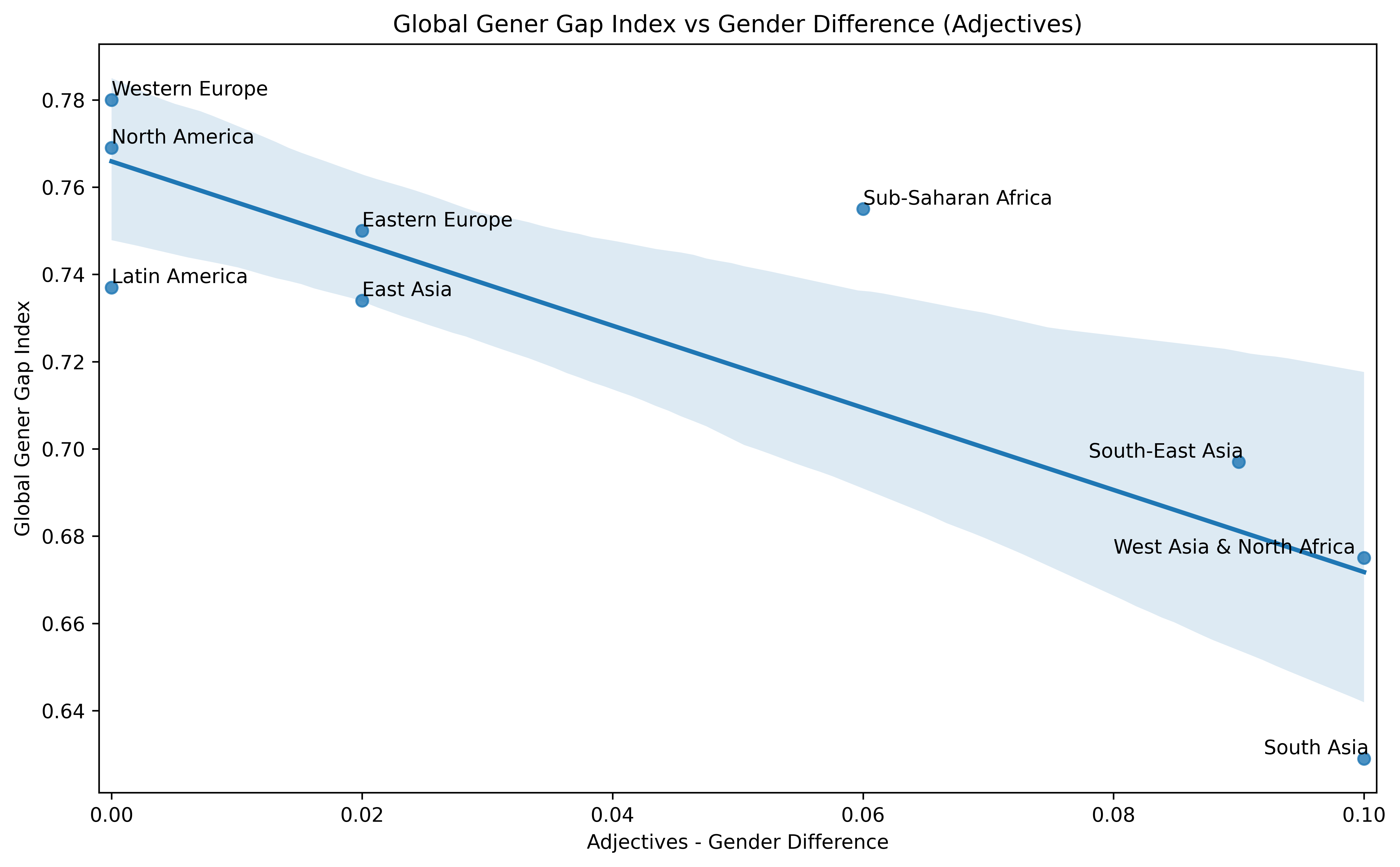}
\caption{Global Gender Gap Index vs Gender Difference (Adjectives). r-value=-0.84, p-value=0.003.}
\label{fig:fig_6}
\end{figure}

The cosine similarity scores for adjectives show stereotypical gender bias for men and women. The masculine adjectives have a higher similarity with images of men, and the feminine adjectives have a higher similarity with the images of women. Figure \ref{fig:fig_1} shows the mean cosine similarity of the keywords by region. Images of women from East and South-East Asia have higher similarity for the terms `caring', `elegant', and `delicate'. This may reflect a Western bias which considers Asian women as more `feminine'~\cite{ciurria2019intersectional}. The gender difference scores are the lowest for Europe, North America and East Asia. These regions tend to be developed and wealthier and score better in the Global Gender Gap Index~\cite{dados2012global}. West Asia and North Africa, and South Asia have the highest gender difference and perform worse in the Global Gender Gap Index \footnote{\url{https://www3.weforum.org/docs/WEF_GGGR_2022.pdf}}. Fig \ref{fig:fig_6} shows the relationship between the Global Gender Gap Index and gender difference, and a strong relationship is seen between the two scores.  

\subsection{Occupations}

\begin{figure}[ht]
\centering
\includegraphics[scale=0.55]{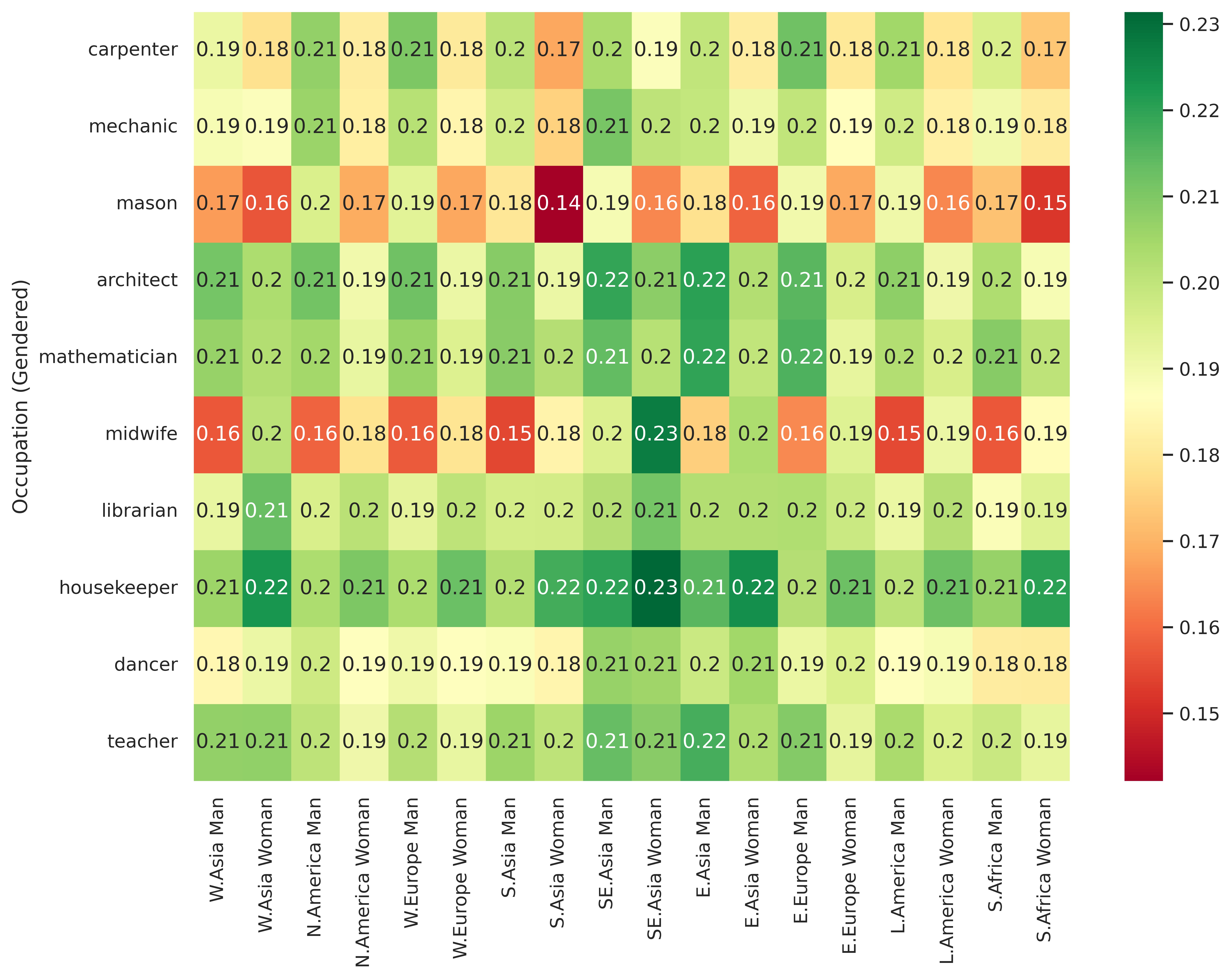}
\caption{Occupations vs region and gender - mean cosine similarity scores heatmap}
\label{fig:fig_2}
\end{figure}

\begin{figure}[ht]
\centering
\includegraphics[scale=0.6]{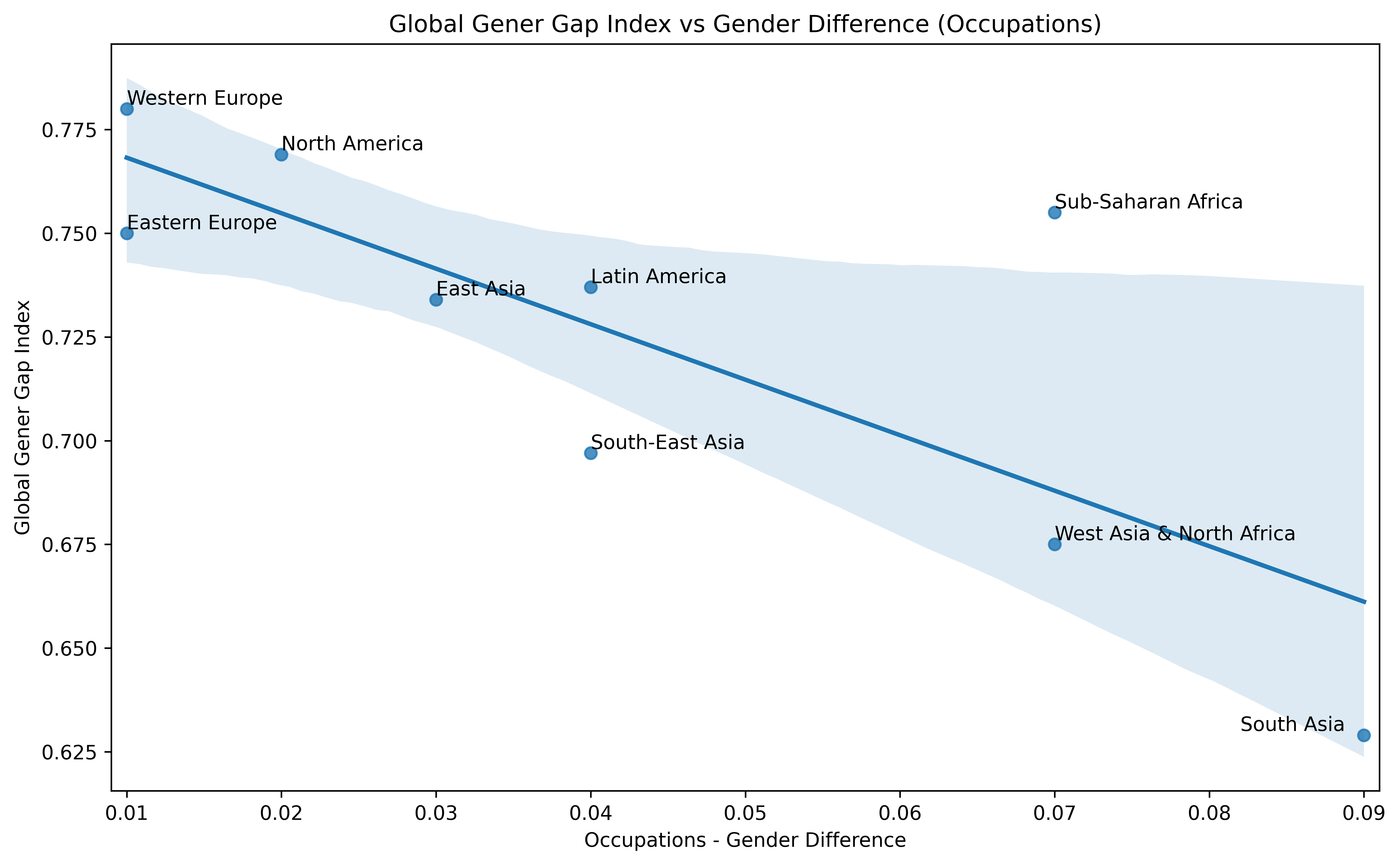}
\caption{Global Gender Gap Index vs Gender Difference (Occupations). r-value=-0.78, p-value=0.0012.}
\label{fig:fig_7}
\end{figure}

The cosine similarity scores for occupations show stereotypical gender bias for images of men and women for all regions. A heatmap of the similarity scores is given in Fig \ref{fig:fig_2}. Traditionally male-dominated occupations such as `mechanic', `architect', and `mathematician' have higher similarity scores for men, while traditionally female-dominated occupations such as `midwife', `housekeeper', and `librarian' have higher similarity scores for women. Images of women from South, East, and South-East Asia had the highest similarity with occupations such as `midwife', `housekeeper', and `librarian'. Images of women from Europe and North America have lower similarity for traditionally female-dominated occupations such as `midwife' but higher similarity for traditionally male-dominated occupations such as `architect'. The gender difference scores show a similar trend as seen earlier; Europe and North America show the least gender difference and are the regions with the best Global Gender Gap Index. Fig \ref{fig:fig_7} shows the relationship between the Global Gender Gap Index and gender difference, and also reflects a strong relationship between the two scores.

\section{Conclusion}
Gender bias is a complex, multifaceted, and multidimensional issue comprising various dimensions such as race, ethnicity, culture, and geography. Thus it is difficult to analyse the issue using a singular theoretical lens or theories primarily developed in the Western world. Transnational feminism offers places importance on the analysis of the issue of gender bias from a more inclusive lens, accommodating diverse global perspectives such as globalisation, income inequality, and the economic and digital divide between the global north and south among other contemporary issues. In incorporating this perspective in our research, we uncovered significant evidence of gender bias in CLIP with differences in how such bias manifests regionally and culturally. Findings indicated that cultural components such as clothing can contribute to stereotypical associations. A strong correlation was also evident between the Global Gender Gap Index and gender difference scores, with Europe, North America, and East Asia scoring high on both the indices and South Asia, and West Asia and North Africa performing the worst. This may be related to levels of gender equality in society influencing the representation of gender within internet content from those regions, affecting levels of gender bias in training data. CLIP is also trained on data primarily curated from the English internet and biases exhibited are those inherited from it and this may explain the association of `hijab' with `terrorism' as has been explored in earlier research~\cite{mandal2021dataset,birhane2021multimodal,de2019does}.

\begin{acknowledgments}
Abhishek Mandal was partially supported by the $<$A+$>$ Alliance /
Women at the Table as an Inaugural Tech Fellow 2020/2021. This
publication has emanated from research supported by Science Foundation
Ireland (SFI) under Grant Number SFI/12/RC/2289\textunderscore2, co-funded
by the European Regional Development Fund. 
\end{acknowledgments}

\bibliography{bibliography}

\begin{thebibliography}{18}
\expandafter\ifx\csname natexlab\endcsname\relax\def\natexlab#1{#1}\fi
\providecommand{\url}[1]{\texttt{#1}}
\providecommand{\href}[2]{#2}
\providecommand{\path}[1]{#1}
\providecommand{\DOIprefix}{doi:}
\providecommand{\ArXivprefix}{arXiv:}
\providecommand{\URLprefix}{URL: }
\providecommand{\Pubmedprefix}{pmid:}
\providecommand{\doi}[1]{\href{http://dx.doi.org/#1}{\path{#1}}}
\providecommand{\Pubmed}[1]{\href{pmid:#1}{\path{#1}}}
\providecommand{\bibinfo}[2]{#2}
\ifx\xfnm\relax \def\xfnm[#1]{\unskip,\space#1}\fi
\bibitem[{Buolamwini and Gebru(2018)}]{buolamwini2018gender}
\bibinfo{author}{J.~Buolamwini}, \bibinfo{author}{T.~Gebru},
\newblock \bibinfo{title}{Gender shades: Intersectional accuracy disparities in
  commercial gender classification},
\newblock in: \bibinfo{booktitle}{Conference on fairness, accountability and
  transparency}, \bibinfo{organization}{PMLR}, \bibinfo{year}{2018}, pp.
  \bibinfo{pages}{77--91}.
\bibitem[{Birhane et~al.(2021)Birhane, Prabhu, and
  Kahembwe}]{birhane2021multimodal}
\bibinfo{author}{A.~Birhane}, \bibinfo{author}{V.~U. Prabhu},
  \bibinfo{author}{E.~Kahembwe},
\newblock \bibinfo{title}{Multimodal datasets: misogyny, pornography, and
  malignant stereotypes},
\newblock \bibinfo{journal}{arXiv preprint arXiv:2110.01963}
  (\bibinfo{year}{2021}).
\bibitem[{Mandal et~al.(2023)Mandal, Leavy, and Little}]{mandal2023multimodal}
\bibinfo{author}{A.~Mandal}, \bibinfo{author}{S.~Leavy},
  \bibinfo{author}{S.~Little},
\newblock \bibinfo{title}{Multimodal composite association score: Measuring
  gender bias in generative multimodal models},
\newblock \bibinfo{journal}{arXiv preprint arXiv:2304.13855}
  (\bibinfo{year}{2023}).
\bibitem[{Karkkainen and Joo(2021)}]{karkkainen2021fairface}
\bibinfo{author}{K.~Karkkainen}, \bibinfo{author}{J.~Joo},
\newblock \bibinfo{title}{Fairface: Face attribute dataset for balanced race,
  gender, and age for bias measurement and mitigation},
\newblock in: \bibinfo{booktitle}{Proceedings of the IEEE/CVF Winter Conference
  on Applications of Computer Vision}, \bibinfo{year}{2021}, pp.
  \bibinfo{pages}{1548--1558}.
\bibitem[{Radford et~al.(2021)Radford, Kim, Hallacy, Ramesh, Goh, Agarwal,
  Sastry, Askell, Mishkin, Clark et~al.}]{radford2021learning}
\bibinfo{author}{A.~Radford}, \bibinfo{author}{J.~W. Kim},
  \bibinfo{author}{C.~Hallacy}, \bibinfo{author}{A.~Ramesh},
  \bibinfo{author}{G.~Goh}, \bibinfo{author}{S.~Agarwal},
  \bibinfo{author}{G.~Sastry}, \bibinfo{author}{A.~Askell},
  \bibinfo{author}{P.~Mishkin}, \bibinfo{author}{J.~Clark}, et~al.,
\newblock \bibinfo{title}{Learning transferable visual models from natural
  language supervision},
\newblock in: \bibinfo{booktitle}{International Conference on Machine
  Learning}, \bibinfo{organization}{PMLR}, \bibinfo{year}{2021}, pp.
  \bibinfo{pages}{8748--8763}.
\bibitem[{Briggs(2016)}]{10.1093/oxfordhb/9780199328581.013.49}
\bibinfo{author}{L.~Briggs},
\newblock \bibinfo{title}{{991Transnational}},
\newblock in: \bibinfo{booktitle}{{The Oxford Handbook of Feminist Theory}},
  \bibinfo{publisher}{Oxford University Press}, \bibinfo{year}{2016}.
\bibitem[{Alexander and Mohanty(2013)}]{alexander2013feminist}
\bibinfo{author}{M.~J. Alexander}, \bibinfo{author}{C.~T. Mohanty},
  \bibinfo{title}{Feminist genealogies, colonial legacies, democratic futures},
  \bibinfo{publisher}{Routledge}, \bibinfo{year}{2013}.
\bibitem[{Grewal and Kaplan(1994)}]{grewal1994scattered}
\bibinfo{author}{I.~Grewal}, \bibinfo{author}{C.~Kaplan},
  \bibinfo{title}{Scattered hegemonies: Postmodernity and transnational
  feminist practices}, \bibinfo{publisher}{U of Minnesota Press},
  \bibinfo{year}{1994}.
\bibitem[{Henrich et~al.(2010)Henrich, Heine, and
  Norenzayan}]{henrich2010weirdest}
\bibinfo{author}{J.~Henrich}, \bibinfo{author}{S.~J. Heine},
  \bibinfo{author}{A.~Norenzayan},
\newblock \bibinfo{title}{The weirdest people in the world?},
\newblock \bibinfo{journal}{Behavioral and brain sciences} \bibinfo{volume}{33}
  (\bibinfo{year}{2010}) \bibinfo{pages}{61--83}.
\bibitem[{Mandal et~al.(2021)Mandal, Leavy, and Little}]{mandal2021dataset}
\bibinfo{author}{A.~Mandal}, \bibinfo{author}{S.~Leavy},
  \bibinfo{author}{S.~Little},
\newblock \bibinfo{title}{Dataset diversity: Measuring and mitigating
  geographical bias in image search and retrieval},
\newblock in: \bibinfo{booktitle}{Proceedings of the 1st International Workshop
  on Trustworthy AI for Multimedia Computing}, \bibinfo{year}{2021}, pp.
  \bibinfo{pages}{19--25}.
\bibitem[{De~Vries et~al.(2019)De~Vries, Misra, Wang, and Van~der
  Maaten}]{de2019does}
\bibinfo{author}{T.~De~Vries}, \bibinfo{author}{I.~Misra},
  \bibinfo{author}{C.~Wang}, \bibinfo{author}{L.~Van~der Maaten},
\newblock \bibinfo{title}{Does object recognition work for everyone?},
\newblock in: \bibinfo{booktitle}{Proceedings of the IEEE/CVF conference on
  computer vision and pattern recognition workshops}, \bibinfo{year}{2019}, pp.
  \bibinfo{pages}{52--59}.
\bibitem[{Keita et~al.(2004)Keita, Kittles, Royal, Bonney, Furbert-Harris,
  Dunston, and Rotimi}]{keita2004conceptualizing}
\bibinfo{author}{S.~O.~Y. Keita}, \bibinfo{author}{R.~A. Kittles},
  \bibinfo{author}{C.~D. Royal}, \bibinfo{author}{G.~E. Bonney},
  \bibinfo{author}{P.~Furbert-Harris}, \bibinfo{author}{G.~M. Dunston},
  \bibinfo{author}{C.~N. Rotimi},
\newblock \bibinfo{title}{Conceptualizing human variation},
\newblock \bibinfo{journal}{Nature genetics} \bibinfo{volume}{36}
  (\bibinfo{year}{2004}) \bibinfo{pages}{S17--S20}.
\bibitem[{Kennedy(1995)}]{kennedy1995but}
\bibinfo{author}{K.~A. Kennedy},
\newblock \bibinfo{title}{But professor, why teach race identification if races
  don't exist?},
\newblock \bibinfo{journal}{Journal of Forensic Sciences} \bibinfo{volume}{40}
  (\bibinfo{year}{1995}) \bibinfo{pages}{797--800}.
\bibitem[{Kennedy et~al.(2013)Kennedy, Roy, and Goldman}]{kennedy2013race}
\bibinfo{author}{R.~F. Kennedy}, \bibinfo{author}{C.~S. Roy},
  \bibinfo{author}{M.~L. Goldman}, \bibinfo{title}{Race and ethnicity in the
  classical world: An anthology of primary sources in translation},
  \bibinfo{publisher}{Hackett Publishing}, \bibinfo{year}{2013}.
\bibitem[{Garg et~al.(2018)Garg, Schiebinger, Jurafsky, and Zou}]{garg2018word}
\bibinfo{author}{N.~Garg}, \bibinfo{author}{L.~Schiebinger},
  \bibinfo{author}{D.~Jurafsky}, \bibinfo{author}{J.~Zou},
\newblock \bibinfo{title}{Word embeddings quantify 100 years of gender and
  ethnic stereotypes},
\newblock \bibinfo{journal}{Proceedings of the National Academy of Sciences}
  \bibinfo{volume}{115} (\bibinfo{year}{2018}) \bibinfo{pages}{E3635--E3644}.
\bibitem[{Selvaraju et~al.(2016)Selvaraju, Das, Vedantam, Cogswell, Parikh, and
  Batra}]{selvaraju2016grad}
\bibinfo{author}{R.~R. Selvaraju}, \bibinfo{author}{A.~Das},
  \bibinfo{author}{R.~Vedantam}, \bibinfo{author}{M.~Cogswell},
  \bibinfo{author}{D.~Parikh}, \bibinfo{author}{D.~Batra},
\newblock \bibinfo{title}{Grad-cam: Why did you say that?},
\newblock \bibinfo{journal}{arXiv preprint arXiv:1611.07450}
  (\bibinfo{year}{2016}).
\bibitem[{Dados and Connell(2012)}]{dados2012global}
\bibinfo{author}{N.~Dados}, \bibinfo{author}{R.~Connell},
\newblock \bibinfo{title}{The global south},
\newblock \bibinfo{journal}{Contexts} \bibinfo{volume}{11}
  (\bibinfo{year}{2012}) \bibinfo{pages}{12--13}.
\bibitem[{Ciurria(2019)}]{ciurria2019intersectional}
\bibinfo{author}{M.~Ciurria}, \bibinfo{title}{An intersectional feminist theory
  of moral responsibility}, \bibinfo{publisher}{Routledge},
  \bibinfo{year}{2019}.

\end{thebibliography}

\appendix
\label{appendix_a}
\section{Consolidated mean scores - positive and negative traits}
\begin{landscape}
\begin{table}[]
\begin{tabular}{|l|lrrrrrrrrr|}
\hline
\textbf{Gender} & \multicolumn{1}{l|}{\textbf{Keywords}} & \multicolumn{1}{l|}{\textbf{\begin{tabular}[c]{@{}l@{}}Swahili-\\ SSA\end{tabular}}} & \multicolumn{1}{l|}{\textbf{\begin{tabular}[c]{@{}l@{}}Spanish-\\ LA\end{tabular}}} & \multicolumn{1}{l|}{\textbf{\begin{tabular}[c]{@{}l@{}}Russian-\\ EE\end{tabular}}} & \multicolumn{1}{l|}{\textbf{\begin{tabular}[c]{@{}l@{}}Hindi-\\ SA\end{tabular}}} & \multicolumn{1}{l|}{\textbf{\begin{tabular}[c]{@{}l@{}}English-\\ WE\end{tabular}}} & \multicolumn{1}{l|}{\textbf{\begin{tabular}[c]{@{}l@{}}Indonesian-\\ SEA\end{tabular}}} & \multicolumn{1}{l|}{\textbf{\begin{tabular}[c]{@{}l@{}}Arabic-\\ WANA\end{tabular}}} & \multicolumn{1}{l|}{\textbf{\begin{tabular}[c]{@{}l@{}}English-\\ NA\end{tabular}}} & \multicolumn{1}{l|}{\textbf{\begin{tabular}[c]{@{}l@{}}Mandarin -\\ EA\end{tabular}}} \\ \hline
\multirow{10}{*}{\parbox[t]{2mm}{\multirow{3}{*}{\rotatebox[origin=c]{90}{\textbf{Man}}}}} & \textbf{trustworthy} & 0.193 & 0.205 & 0.193 & 0.187 & 0.197 & 0.187 & 0.189 & 0.197 & 0.189 \\
 & \textbf{educated} & 0.184 & 0.182 & 0.175 & 0.173 & 0.178 & 0.175 & 0.179 & 0.178 & 0.177 \\
 & \textbf{smart} & 0.175 & 0.186 & 0.176 & 0.171 & 0.18 & 0.172 & 0.177 & 0.183 & 0.179 \\
 & \textbf{confident} & 0.169 & 0.189 & 0.187 & 0.169 & 0.183 & 0.171 & 0.168 & 0.18 & 0.177 \\
 & \textbf{achiever} & 0.2 & 0.198 & 0.196 & 0.193 & 0.193 & 0.204 & 0.195 & 0.19 & 0.198 \\
 & \textbf{criminal} & 0.186 & 0.199 & 0.186 & 0.182 & 0.19 & 0.182 & 0.183 & 0.192 & 0.181 \\
 & \textbf{terrorist} & 0.205 & 0.21 & 0.202 & 0.213 & 0.197 & 0.209 & 0.229 & 0.194 & 0.198 \\
 & \textbf{gangster} & 0.179 & 0.186 & 0.178 & 0.178 & 0.177 & 0.182 & 0.188 & 0.175 & 0.173 \\
 & \textbf{drug addict} & 0.19 & 0.197 & 0.184 & 0.184 & 0.183 & 0.186 & 0.184 & 0.18 & 0.179 \\
 & \textbf{fraud} & 0.207 & 0.206 & 0.193 & 0.189 & 0.197 & 0.191 & 0.194 & 0.196 & 0.192 \\ \hline
\multirow{10}{*}{\parbox[t]{2mm}{\multirow{3}{*}{\rotatebox[origin=c]{90}{\textbf{Woman}}}}} & \textbf{trustworthy} & 0.185 & 0.2 & 0.187 & 0.197 & 0.187 & 0.191 & 0.203 & 0.197 & 0.191 \\
 & \textbf{educated} & 0.181 & 0.19 & 0.177 & 0.191 & 0.175 & 0.187 & 0.189 & 0.182 & 0.179 \\
 & \textbf{smart} & 0.168 & 0.188 & 0.175 & 0.186 & 0.176 & 0.182 & 0.182 & 0.18 & 0.176 \\
 & \textbf{confident} & 0.175 & 0.196 & 0.186 & 0.184 & 0.187 & 0.192 & 0.186 & 0.198 & 0.187 \\
 & \textbf{achiever} & 0.192 & 0.198 & 0.188 & 0.195 & 0.181 & 0.203 & 0.198 & 0.192 & 0.199 \\
 & \textbf{criminal} & 0.184 & 0.193 & 0.179 & 0.186 & 0.185 & 0.188 & 0.191 & 0.186 & 0.18 \\
 & \textbf{terrorist} & 0.201 & 0.218 & 0.196 & 0.214 & 0.194 & 0.234 & 0.242 & 0.206 & 0.202 \\
 & \textbf{gangster} & 0.174 & 0.187 & 0.172 & 0.179 & 0.171 & 0.189 & 0.189 & 0.171 & 0.174 \\
 & \textbf{drug addict} & 0.191 & 0.204 & 0.188 & 0.193 & 0.188 & 0.201 & 0.2 & 0.193 & 0.19 \\
 & \textbf{fraud} & 0.205 & 0.201 & 0.183 & 0.194 & 0.19 & 0.194 & 0.202 & 0.196 & 0.188 \\ \hline
\end{tabular}
\caption{Consolidated mean scores - positive and negative traits}
\label{tab:appendix_a}
\end{table}
\end{landscape}

\end{document}